\documentclass[12pt,a4paper]{article}
\usepackage{jcappub}

\usepackage{amsmath,amssymb}

\def\bx{{\boldsymbol x}}
\def\bk{{\boldsymbol k}}
\def\cF{{\cal F}}
\def\eff{{\rm eff}}
\def\ri{{\rm i}}

\title{Magnetogenesis by non-minimal coupling to gravity in the Starobinsky inflationary model}

\author[a]{Oleg Savchenko}
\author[a,b]{and Yuri Shtanov}

\affiliation[a]{Department of Physics, Taras Shevchenko National University,  03022 Kiev, Ukraine} %
\affiliation[b]{Bogolyubov Institute for Theoretical Physics,  03143 Kiev, Ukraine} %

\emailAdd{savchenkooleg42@gmail.com}
\emailAdd{shtanov@bitp.kiev.ua}

\abstract{The $R^2$ term in the Starobinsky inflationary model can be regarded as a leading quantum correction to the gravitational effective action.  We assume that parity-preserving and parity-violating (axial) non-minimal couplings between curvature and electromagnetic field are also present in the effective action.  In the Einstein frame, they turn into non-trivial couplings of the scalaron and curvature to the electromagnetic field.  We make an assessment of inflationary magnetogenesis in this model.  In the case of parity-preserving couplings, amplification of magnetic field is negligibly small. In the case of axial couplings, magnetogenesis is hampered by strong back-reaction on the inflationary process, resulting in possible amplification of magnetic field at most by the factor $10^5$ relative to its vacuum fluctuations.}


\begin{document} 
\maketitle
\flushbottom


\section{Introduction}

Magnetic fields are present in our universe on a broad range of spatial scales.  Spiral galaxies similar to Milky Way host regular magnetic fields of the order of $\mu$G, while distant galaxies exhibit fields of the order of $100~\mu$G \cite{Bernet:2008qp, Wolfe:2008nk}. There is a strong evidence for the presence of magnetic fields in intergalactic medium, including voids \cite{Tavecchio:2010mk, Ando:2010rb, Neronov:1900zz, Dolag:2010}, with strengths $\gtrsim 10^{-16}$~G\@.  All this suggests a cosmological origin of magnetic fields, which are subsequently amplified in galaxies, probably by the dynamo mechanism (see reviews \cite{Grasso:2000wj, Widrow:2002ud, Kandus:2010nw, Durrer:2013pga, Subramanian:2015lua}).

Various mechanisms of the cosmological origin of magnetic fields have been under consideration in the literature (for reviews, see \cite{Grasso:2000wj, Widrow:2002ud, Kandus:2010nw, Durrer:2013pga, Subramanian:2015lua}).  This paper will be concerned with inflationary scenario of magnetogenesis. Two of its classical versions are based on coupling either the inflaton field $\phi$ or the metric curvature to the electromagnetic field in order to violate the conformal invariance of the latter. In the seminal paper \cite{Turner:1987bw},  Turner and Widrow considered gravitational couplings of the (symbolic) type $R A^2$ and $R F^2$, while Ratra \cite{Ratra:1991bn} introduced coupling of the form $e^{\alpha \phi} F^2$ with constant $\alpha$.  In the subsequent development of these ideas, numerous generalizations of the form $ f( \phi ) F^2$ and axial couplings of the form $f ( \phi ) F \tilde F$ as well as their combinations $f(\phi, R) F^2$ and $f(\phi, R) F\tilde F$ were under investigation (see \cite{Durrer:2013pga, Subramanian:2015lua} for recent reviews). 

Such inflaton couplings to electromagnetic field are usually introduced ad hoc. On the other hand, couplings to metric curvature, as noted already in \cite{Turner:1987bw}, are naturally expected due to one-loop vacuum polarization in curved space-time.  In this paper, we point out that the vacuum-polarization corrections that are used in the curvature-based models of inflation also naturally generate inflaton coupling to the electromagnetic field.  The seminal inflationary model of this kind is the Starobinsky model \cite{Starobinsky:1980te}, for which we illustrate this idea.  We consider parity-preserving couplings of the form $R F^2$ as well as parity-violating (axial) couplings $R F \tilde F$ as present in the original (Jordan) frame along with the lowest-order correction to the gravitational action proportional to $R^2$.  In the Einstein frame, such terms naturally produce additional non-trivial couplings between the scalaron and the electromagnetic field.  

We then make an assessment of inflationary magnetogenesis in the arising model.  In the case of parity-preserving couplings, magnetic field of considerable strength might be generated only by fine tuning of the coupling constants, while, in the axial case, generation of magnetic field is easier.  However, in both cases, magnetogenesis is hampered by strong back-reaction on the inflationary process.

Our paper is structured as follows. Section~\ref{sec:Starobinsky} describes the Starobinsky model and its inflationary regime.   Section~\ref{sec:parity} introduces parity-preserving couplings to curvature and explores magnetogenesis in the arising model, while section~\ref{sec:axial} does this for parity-violating (axial) couplings.  Section~\ref{sec:both} briefly considers the presence of both types of couplings. We discuss our results in section~\ref{sec:discuss}.

\section{The Starobinsky inflation}
\label{sec:Starobinsky}

\subsection{$f(R)$ gravity and conformal transformation}

In this section, we recall the basic properties of $f (R)$ gravity. We work in the metric signature $(+,-,-,-)$ and in the natural units $\hbar = c = 1$.   The general $f(R)$ action for gravity is
\begin{equation}
\label{Jordan}
S [g_{\mu \nu}] = - \frac{M^2}{3} \int d^4 x \sqrt{- g} f (R) \, ,
\end{equation}
where $M$ is a coupling constant of dimension mass, and the numerical factor is chosen for future convenience.  This theory is conformally equivalent to the usual Einstein--Hilbert gravity plus a self-interacting scalar field (scalaron). Indeed, by introducing an auxiliary dimensionless scalar $\chi$, one can write action (\ref{Jordan}) in the (classically) equivalent form
\begin{equation}\label{Jordan1}
S [g_{\mu \nu}, \chi] = - \frac{M^2}{3} \int d^4x\sqrt{-g} \left[ \chi R - s (\chi) \right]
\end{equation}
with the appropriate function $s (\chi)$.  Then, performing the conformal transformation of the metric $g_{\mu\nu} \to \chi^{-1} g_{\mu\nu}$, one transforms theory \eqref{Jordan1} to
\begin{equation}
S [g_{\mu \nu}, \chi] = \frac{M^2}{3} \int d^4x\sqrt{-g} \left[ -R + \frac{3}{2} \frac{\nabla_\mu \chi\, \nabla^\mu \chi}{\chi^2} + \frac{r(\chi)}{\chi} - \frac{f\left( r(\chi) \right)}{\chi^2} \right] \, ,
\end{equation}
where $r (\chi)$ is the solution of the equation $\chi = f' \left( r \right)$ with respect to $r$. 
After that, changing the scalar-field variable according to 
\begin{equation}\label{chi}
\chi (\phi) = e^{\phi / M} \, ,	
\end{equation}
one  obtains a theory of canonical scalar $\phi$ minimally coupled to the Einstein gravity,
\begin{equation} \label{Einstein}
S [g_{\mu \nu}, \phi] = - \frac{M^2}{3}\int d^4x\sqrt{-g}R +  \int d^4x \sqrt{-g} \left[ \frac12 \nabla_\mu \phi \nabla^\mu \phi 
- V (\phi) \right] \, ,
\end{equation}
in which the scalar-field potential is given by
\begin{equation}
V (\phi) = \frac{M^2}{3} \left[ \chi^{-2} (\phi)  f \left( r \left( \chi (\phi) \right) \right) - \chi^{-1} (\phi) r \left( \chi (\phi) \right) \right] \, .
\end{equation}

Action (\ref{Jordan1}), and sometimes also (\ref{Jordan}), is said to be written in the `Jordan frame' of field variables, while (\ref{Einstein}) is referred to as the corresponding action in the `Einstein frame.'  The constant $M$ is then related to the gravitational constant $G$ as  $M^2 = 3 / 16 \pi G$ in the natural units $\hbar = c = 1$.

\subsection{The Starobinsky model}

The Starobinsky model \cite{Starobinsky:1980te} is a special case of (\ref{Jordan}) with the action
\begin{equation} \label{Jordan-S}
S [g_{\mu \nu}] = - \frac{M^2}{3}\int d^4x\sqrt{-g} \left( R - \frac{R^2}{6 m^2} \right) \, .
\end{equation}
The second term in this action can be regarded as stemming from the lowest-order quantum correction to the effective action for gravity.  For this model, the transformations of the previous section can be done explicitly:
\begin{equation}
r (\chi) = 3 m^2(1-\chi)\, , 
\end{equation}
\begin{equation} \label{pot}
V (\phi) = \frac12 m^2 M^2 \left( 1 - e^{- \phi / M} \right)^2 \, ,
\end{equation}
\begin{equation} \label{Einstein-S}
S [g_{\mu \nu}, \phi]= - \frac{M^2}{3} \int d^4x \sqrt{-g} R + \int d^4x \sqrt{-g} \left[ \frac{1}{2}(\nabla\phi)^2 - V(\phi) \right] \, .
\end{equation}
The scalar-field potential (\ref{pot}) has minimum at $\phi = 0$, with $m$ being its mass: $m^2 = V''(0)$.  Inflation based on this model is in excellent agreement with current observations \cite{Akrami:2018odb}, and the observed amplitude of the primordial power spectrum of cosmological perturbations fixes the inflaton mass to be 
\begin{equation}\label{mass}
m \approx 10^{-5} M \, .	
\end{equation}

\subsection{Inflationary regime}

 In describing the spatially flat Friedmann metric, we work with the physical time $t$ as well as with the conformal time $\eta$: 
\begin{equation}
ds^2 = d t^2 - a^2 (t) d\bx^2 = a^2 (\eta) \left( d\eta^2 - d\bx^2 \right) = a^2(\eta)\eta_{\mu \nu}dx^{\mu}dx^{\nu} \, .
\end{equation}
Derivatives with respect to $t$ are denoted by an overdot, while those with respect to $\eta$ by a prime.

In the inflationary regime, the scalar field is slowly rolling down its potential from positive values of $\phi$ towards zero, so that
\begin{equation} \label{slow-roll}
	\dot \phi \approx - \frac{V' (\phi)}{3 H} \, , \qquad H^2 \approx \frac{V (\phi)}{2 M^2} \, , \qquad \dot \phi^2 \ll V (\phi) \, , \qquad a(\eta) \approx -\frac{1}{H \eta}\, ,
\end{equation}
where $H = \dot a / a$.  The inflationary and slow-roll parameters are given by
\begin{align}
	\epsilon_V &\equiv - \frac{\dot H}{H^2} \approx \frac{M^2}{3} \left[ \frac{V' (\phi) }{V (\phi)} \right]^2 = \frac43 \left( e^{\phi / M} -1 \right)^{-2} \, , \\
	\eta_V &\equiv \epsilon_V - \frac{\ddot \phi}{H \dot \phi} \approx \frac{2 M^2}{3} \frac{V'' (\phi)}{V (\phi)} = \frac{4 \left( 2 - e^{\phi/M} \right)}{3 \left( e^{\phi/M} - 1 \right)^2 } \, ,
\end{align}
and the inflationary and slow-roll conditions $\epsilon_V \ll 1$ and $\eta_V \ll 1$ are both satisfied if $e^{\phi / M} \gg 1$.  In this regime, we have 
\begin{equation}
	H \approx \frac{m}{2} \left( 1 - e^{- \phi/ M} \right)	\, ,
\end{equation}
and the slow-roll dynamics (\ref{slow-roll}) is described by
\begin{equation}
	\dot \phi \approx - \frac23 m M e^{- \phi / M} \, .
\end{equation}
These expressions allow one to relate the scalar field to the scale factor: $ e^{\phi / M} = {\rm const} - \frac43 \log a$.  The integration constant is determined from the conventional condition $\epsilon_V = 1$ at the end of inflation (labeled by index `f').  At this moment, we have $e^{\phi / M} \approx 2$, so that
\begin{equation}\label{ephi}
	e^{\phi / M} \approx \frac43 \log \frac{a_{\rm f}}{a} + 2 \, .
\end{equation}

\section{Parity-preserving couplings}
\label{sec:parity}

\subsection{Lagrangian}

The usual Maxwell action
\begin{equation} \label{Maxwell}
S_0  = - \frac14 \int d^4x\sqrt{-g} F_{\mu \nu} F^{\mu \nu} 
\end{equation}
is invariant with respect to conformal transformations of the metric, and, therefore, does not change its form when  proceeding from the Jordan to the Einstein frame.  

Just as one can consider the second term in action (\ref{Jordan-S}) as generated by quantum corrections, one can expect generation of non-minimal couplings between gravity and electromagnetic field that will violate the conformal invariance.  Non-minimal parity-preserving couplings of the smallest possible dimension six have the form
\begin{equation} \label{Snorm}
S_{\rm int}  =  \int d^4x\sqrt{-g} \left[ \kappa_1 R F_{\mu \nu} 
F^{\mu \nu}  + \kappa_2 \widetilde R_{\mu \nu}  F^\mu{}_{\sigma} 
F^{\sigma\nu} + \kappa_3 C_{\lambda \rho \mu \nu}\, F^{\lambda \rho} F^{\mu \nu} \right] \equiv S_1 + S_2 + S_3 \, ,
\end{equation}
where $ \widetilde R_{\mu\nu} \equiv R_{\mu\nu} - \frac14 g_{\mu\nu} R$ is the trace-free part of the Ricci tensor, 
$C_{\lambda \rho \mu \nu}$ is the Weyl tensor, and $\kappa_i$, $i = 1, 2, 3$, are  constants of dimension inverse mass squared.  

Such couplings to electromagnetism can actually arise from similar gauge-invariant couplings to the hypercharge and weak-isospin field strengths.  We note, however, that, during inflation, the Higgs scalar field exhibits quantum fluctuations of magnitude $\sim H$ on super-Hubble spatial scales, where $H$ is the Hubble parameter, so that the only massless gauge field in the electroweak sector is the electromagnetic field.  Hence, we study coupling to this field only as most relevant for magnetogenesis.

After proceeding to the Einstein frame by the conformal transformation $g_{\mu \nu} \to e^{-\phi / M} g_{\mu\nu}$, the components of this action transform as follows:
\begin{align}
S_1 &= \kappa_1 \int d^4x\sqrt{-g}\, e^{\phi / M} \left[ R + \frac{3}{M}\Box\phi - \frac{3}{2 M^2}(\nabla\phi)^2\right] F_{\mu \nu} F^{\mu \nu} \, , \label{S1} \\
S_2 &= \kappa_2 \int d^4x\sqrt{-g}\, e^{\phi / M}  \left[ \widetilde R_{\mu \nu} + \frac{1}{M}\left( \nabla_{\mu} \nabla_{\nu}\phi - \frac14 g_{\mu \nu} \Box \phi \right) \right. \nonumber \\
 &\qquad \left. {} + \frac{1}{2 M^2} \left( \nabla_{\mu}\phi \nabla_{\nu} \phi - \frac14 g_{\mu \nu} (\nabla \phi)^2  \right) \right] F^{\mu \lambda} F_{\lambda}{}^{\nu} \, , \label{S2} \\
S_3 &= \kappa_3 \int d^4x\sqrt{-g}\, e^{\phi / M} \, C_{\lambda \rho \mu \nu}\, F^{\lambda \rho} F^{\mu \nu} \, , \label{S3}
\end{align}
where all covariant operations are now performed using the transformed metric.  

We note the appearance of the scalar field $\phi$ in each of the terms (\ref{S1})--(\ref{S3}).  Thus, in the Einstein frame of field variables, we have arrived at a hybrid model of modified electrodynamics combining the curvature couplings as in \cite{Turner:1987bw} and a non-trivial inflaton coupling, including the exponential factor similar to that of \cite{Ratra:1991bn}. (The Starobinsky model with exponential coupling of the inflaton to electromagnetic field was under consideration in  \cite{Vilchinskii:2017qul}.)  

In the case of Friedmannian universe, the spacetime metric is conformally flat, and, therefore, its Weyl tensor vanishes.  Thus, the presence of part (\ref{S3}) in the action does not contribute to the  equations of motion of the scalaron or electromagnetic field.  However, the presence of this term in general is important for the considerations of back-reaction since it will contribute to the stress--energy tensor of the magnetic field.

We are going to treat the electromagnetic field as a test field, ignoring its impact (back-reaction) on the spacetime metric or on the scalaron $\phi$ (the validity of this approximation will be investigated afterwards).  In other words, we will work in the linear approximation for the electromagnetic field.  In this case, the effective action for the electromagnetic field can be further reduced by taking into account the Einstein equation,
\begin{equation}
R_{\mu\nu} - \frac12 g_{\mu\nu} R = \frac{3}{2 M^2} \left[ \nabla_\mu \phi \nabla_\nu \phi - g_{\mu\nu} \left( \frac12 \left( \nabla \phi \right)^2 - V (\phi ) \right) + T_{\mu\nu} \right] \, ,
\end{equation}
and the evolution equation for the scalar field,
\begin{equation} \label{eq-phi}
\Box \phi + V' (\phi) = \frac{\delta S_{\rm m}}{\delta \phi} \, ,
\end{equation}
where $T_{\mu\nu} = 2 ( - g)^{-1/2} \delta S_{\rm m}/\delta g^{\mu\nu}$ is the stress-energy tensor for the rest of matter, and $S_{\rm m}$ is the matter Lagrangian (including the matter coupling to the scalaron $\phi$).\footnote{We write the contributions from the rest of matter just for completeness; in calculations they will always be neglected.}  Using these equations in (\ref{S1}) and (\ref{S2}), we obtain the corresponding effective actions:
\begin{align}
S_1 &= - 3 \kappa_1 \int d^4x\sqrt{-g}\, e^{\phi / M} \left( \frac{1}{M} \left[ V' (\phi) - \frac{\delta S_{\rm m}}{\delta \phi} \right] 
+ \frac{1}{2 M^2} \left[ 4 V (\phi) + T \right] \right) F_{\mu \nu} F^{\mu \nu} \, , \label{S-1} \\
S_2 &= \kappa_2 \int d^4x\sqrt{-g}\, e^{\phi / M}  \left[  \frac{1}{M} \left( \nabla_{\mu} \nabla_{\nu}\phi - \frac14 g_{\mu \nu} \Box \phi \right) \right. \nonumber \\
 &\qquad \left. {} + \frac{2}{M^2} \left( \nabla_{\mu}\phi \nabla_{\nu} \phi - \frac14 g_{\mu \nu} (\nabla \phi)^2  + \frac34 \widetilde T_{\mu\nu} \right) \right] F^{\mu \lambda} F_{\lambda}{}^{\nu} \, ,\label{S-2} 
\end{align}
where $T$ is the trace, and $\widetilde T_{\mu\nu}$ is the trace-free part of $T_{\mu\nu}$.

We decompose $A_{\mu}$ into its transverse and longitudinal parts: $A_{\mu}=(A_0, A_i)$, $A_i=A_i^T+\partial_i \chi$, with $\partial_i A_i^T=0$. The variables $A_0$ and $\chi$ are then eliminated from the action, and the usual Maxwell action (\ref{Maxwell}) on the cosmological background becomes
\begin{equation} \label{S_0}
S_0 = \frac12 \int d^4x \left( A_i^{T\prime} A_i^{T\prime} + A_i^T \Delta A_i^T \right) \, ,
\end{equation}
where $d^4x\equiv d\eta d^3\bx$, and the Laplacian $\Delta$ is calculated with respect to the Euclidean metric in the $\bx$ space.  Here and below, the prime denotes the derivative with respect to the conformal time, except in the expression $V' (\phi)$, where it explicitly denotes the derivative with respect to $\phi$. 

For homogeneous field $\phi = \phi(\eta)$ and $T_{\mu\nu}$ describing homogeneous isotropic matter distribution with density $\rho$ and pressure $p$, actions (\ref{S-1}) and (\ref{S-2}) take the following form:
\begin{align}
S_1 &= 3 \kappa_1 \int d^4x\, e^{\phi / M} \left( \frac{2}{M} \left[ V' (\phi) - \frac{\delta S_{\rm m}}{\delta \phi} \right] \right. \nonumber \\ &\qquad \left. {} + \frac{1}{M^2} \left[ 4 V (\phi) + \rho - 3 p \right] \right) \left( A_i^{T\prime} A_i^{T\prime} + A_i^T  \Delta A_i^T\right)  \, , \label{S_1} \\
S_2 &= \kappa_2 \int d^4x\, e^{\phi / M}  \left( \frac{1}{2 M}\left[ 4 H \dot \phi + V' (\phi) - \frac{\delta S_{\rm m}}{\delta \phi} \right] \right. \nonumber \\ &\qquad \left. {} 
- \frac{1}{M^2} \left[ \dot \phi^2 + \frac34 (\rho + p) \right] \right) 
\left( A_i^T  \Delta A_i^T - A_i^{T\prime} A_i^{T\prime} \right) \, , \label{S_2} 
\end{align}
where an overdot denotes the derivative with respect to the cosmological time $t = \int a d \eta$, and $H = \dot a / a$.   In obtaining these equations, we have taken into account (\ref{eq-phi}) to eliminate the second time derivative of the scalaron.

Combining (\ref{S_0})--(\ref{S_2}), we arrive at the action for electromagnetic field in the form
\begin{equation}\label{Sfin}
S  =S_0  + S_1 + S_2 =  \frac12 \int d^4x\, \left[ I_+^2 (\eta) A_i^{T\prime} A_i^{T\prime} + I_-^2 (\eta) A_i^T \Delta A_i^T \right] \, ,
\end{equation}
where 
\begin{align} \label{Ipm}
I_\pm^2 &= 1 + 6 \kappa_1 e^{\phi / M} \left( \frac{2}{M} \left[ V' (\phi) - \frac{\delta S_{\rm m}}{\delta \phi} \right] + \frac{1}{M^2} \left[ 4 V (\phi) + \rho - 3 p \right] \right) \nonumber \\ &\quad {} \mp \kappa_2 e^{\phi / M} \left( \frac{1}{M}\left[ 4 H \dot \phi + V' (\phi) - \frac{\delta S_{\rm m}}{\delta \phi} \right] 
 - \frac{2}{M^2} \left[ \dot \phi^2 + \frac34 (\rho + p) \right] \right) \, .
\end{align}
The quantities $I_\pm^2 (\eta)$ will be assumed to be always positive, in order to avoid instability.

Proceeding to the new variables
\begin{equation} \label{vi}
v_i = I_+ A_i^T \, ,
\end{equation}
and integrating by part in (\ref{Sfin}), we obtain the action
\begin{equation}
S  =  \frac12 \int d^4x\, \left[ v'_i v'_i + \frac{I''_+}{I_+} v_i v_i + \frac{I_-^2}{I_+^2} v_i \Delta v_i \right] \, ,
\end{equation}
and the equation of motion for the Fourier mode of $v_i$ with wavenumber $k$\,:
\begin{equation} \label{eq-v}
v''_i + \left( \frac{I_-^2}{I_+^2} k^2 - \frac{I''_+}{I_+} \right) v_i = 0 \, . 
\end{equation}

For small wavenumbers $k$, namely, for 
\begin{equation}\label{sk}
\frac{I_-^2}{I_+^2} k^2 \ll \left| \frac{I''_+}{ I^{}_+} \right| \, , 	
\end{equation}
the first term in the parentheses of (\ref{eq-v}) can be neglected, and the remaining equation has general solution in the form
\begin{equation}\label{gen}
v_i = C_1 I_+ + C_2 I_+ \int \frac{d \eta}{I_+^2} \, ,
\end{equation}
where $C_1$ and $C_2$ are integration constants.

The combination that enters (\ref{Ipm}) is
\begin{equation} \label{comb}
	e^{\phi/M} \left[ \frac{1}{M} V' (\phi) + \frac{2}{M^2} V (\phi) \right] = m^2 \left( e^{\phi/M} - 1 \right) \approx m^2 \left( \frac43 \log \frac{a_{\rm f}}{a} + 1 \right) \, ,
\end{equation}
where we have used (\ref{ephi}) for the inflationary regime. During inflation, the relevant terms in (\ref{Ipm}) in the slow-roll approximation then give 
\begin{align} \label{Ipm1}
	I_\pm^2 &= 1 + 12 \kappa_1 m^2 \left( e^{\phi/M} - 1 \right)  \pm \frac{\kappa_2 m^2}{3} \left( 1 +  \frac53 e^{- \phi/M} \right) \nonumber \\ &\approx 1 + 12 \kappa_1 m^2 \left( \frac43 \log \frac{a_{\rm f}}{a} + 1 \right) \pm \frac{\kappa_2 m^2}{3} \left( 1 + \frac{5}{4 \log \frac{a_{\rm f}}{a} + 6 }  \right) \, .
\end{align}

\subsection{When $S_1$ dominates}
\label{sec:S1}

In the case of negligible or zero constant $\kappa_2$, the appropriately normalized solution of (\ref{eq-v}) after the Hubble-radius crossing is given by
\begin{equation}
v_i \simeq {\cal O} (1) \frac{I}{I_k \sqrt{k}} \, ,
\end{equation} 
where $I \equiv I_+ = I_-$, and $I_k$ is its value at the Hubble-radius crossing, where $k \simeq aH$.   The vector potential mode in this case is 
\begin{equation}
A_i = \frac{v_i}{I} \simeq \frac{{\cal O} (1)}{I_k \sqrt{k}} \, ,
\end{equation}
and since $I_k \gtrsim 1$, the fluctuations on super-Hubble modes during and after inflation remain to be of the same order of magnitude as in the unmodified electrodynamics.

\subsection{When $S_2$ dominates}

Consider the simplest case of negligible or zero constant $\kappa_1$, so that $S_2$ is the dominating contribution to the action.  There are two regimes of solutions in this case.  For 
\begin{equation}\label{gk}
\frac{I_-^2}{I_+^2} k^2 \gg \left| \frac{I''_+}{ I^{}_+} \right| \, , 
\end{equation}
the second term in the parentheses of (\ref{eq-v}) can be neglected, and  the approximate quantum-normalized positive-frequency solution of (\ref{eq-v}) during inflation is given by 
\begin{equation}\label{vinorm}
v_i = \frac{\varepsilon_i}{\sqrt{2 k_\eff}} e^{- \ri k_\eff \eta} \, ,
\end{equation}
where $\varepsilon_i$ is the normalized polarization vector, and
\begin{equation}
k_\eff = k \sqrt{\frac{3 - \kappa_2 m^2}{3 +\kappa_2 m^2}} \, .
\end{equation}

One can observe the `scaling' renormalisation of the amplitude (and of the speed of light) compared to the standard vacuum value.  The quantity $I_-^2$ is assumed to be always positive (to ensure stability of the electromagnetic field); from (\ref{Ipm1}) we then observe that this requires $\kappa_2 m^2 \lesssim 1$.

Soon after crossing the high-frequency threshold, 
condition (\ref{sk}) becomes valid, and the evolution is given by $v_i \propto I_+$.  Assuming that the subsequent post-inflationary evolution does not significantly influence the relevant electromagnetic modes, we may expect the outcome fluctuations of the magnetic field on the spatial scale corresponding to $k$ to be given by
\begin{equation} \label{k2}
\delta_B (k) \simeq \left( \frac{k}{k_\eff} \right)^{1/2} \delta^{\rm vac}_B (k) = \left( \frac{3 + \kappa_2 m^2}{3 - \kappa_2 m^2} \right)^{1 / 4} \delta^{\rm vac}_B (k) \, .
\end{equation}
Since $\kappa_2 m^2 \lesssim 1$, the amplification factor in this equation is approximately equal to unity.

Thus, we conclude that our model with couplings (\ref{Snorm}) cannot lead to successful magnetogenesis during inflation.

\subsection{Preheating}

The preheating stage is defined here to be the stage immediately following inflation at which the matter effects on the electromagnetic field still can be neglected.  This concerns both the stress-energy tensor of matter in the evolution equations and bulk effects such as electric conductivity of plasma.

At the preheating stage, the scalaron field is oscillating with frequency $m \gg H $ around the minimum of its potential with gradually decreasing amplitude, so that we may also use the approximation $\phi / M \ll 1$.  In this case, from (\ref{Ipm}) we have
\begin{equation}
I_\pm^2 \approx 1 + \left( 12 \kappa_1 \mp \kappa_2 \right)	m^2 \frac{\phi}{M} + 3 \left( \kappa_1 \pm \frac14 \kappa_2 \right) m^2 \frac{\phi_0^2}{M^2}\, ,
\end{equation}
where $\phi_0$ is the amplitude of scalaron oscillations.  In obtaining the last equation, we have averaged (\ref{Ipm}) over one period of oscillations and discarded the oscillatory terms of higher order in $\phi / M$.  We have assumed $I_\pm^2$ to be nonnegative, which must be ensured by sufficient smallness of the dimensionless parameters $\kappa_1 m^2$ and $\kappa_2 m^2$.  In this case, from equation (\ref{eq-v}) we can expect resonant amplification of the modes with $k/a \approx m/2$.  The relevant modes with much lower values of $k$ will be described by solution (\ref{gen}), and will remain non-amplified.

\section{Axial couplings}
\label{sec:axial}

\subsection{Lagrangian}

One can just as well consider terms in the original effective action in the Jordan frame that do not respect the parity symmetry:
\begin{equation}\label{Sax}
S_{\rm int}^{\rm a}  = - \int d^4x\sqrt{-g} \left[ \chi_1 R F_{\mu \nu} \tilde F^{\mu \nu}  + \chi_2 C_{\lambda \rho \mu \nu}\, F^{\lambda \rho} \tilde F^{\mu \nu} \right] \equiv S_{\rm a1} + S_{\rm a2} \, , 
\end{equation}
where $\tilde F^{\mu \nu} \equiv \frac{1}{2} \epsilon^{\mu \nu \rho \sigma} F_{\rho \sigma}$ is the dual of the electromagnetic field stress tensor.  Note that\footnote{This property is most easily established by using the spinor embedding of the tensor algebra.} $F^\mu{}_{\sigma} \tilde F^{\sigma\nu} \propto g^{\mu\nu}$; for this reason the term $\widetilde R_{\mu \nu} F^\mu{}_{\sigma} \tilde F^{\sigma\nu}$ is not present in the Lagrangian. The term $S_{\rm a2}$ is cosmologically irrelevant as it vanishes on the Friedmann background.   Thus, the structure of the axial interaction is particularly simple, and, after proceeding to the Einstein frame by the conformal transformation $g_{\mu \nu} \to e^{-\phi / M} g_{\mu\nu}$ and using the equation of motion for the scalar field, we transform the only relevant component of the action as follows:
\begin{align}
S_{\rm a1} &= - \chi_1 \int d^4x\sqrt{-g}\, e^{\phi / M} \left[ R + \frac{3}{M}\Box\phi - \frac{3}{2 M^2}(\nabla\phi)^2\right] F_{\mu \nu} \tilde F^{\mu \nu} \nonumber \\ &= 3 \chi_1 \int d^4x\sqrt{-g}\, e^{\phi / M} \left( \frac{1}{M} \left[ V' (\phi) - \frac{\delta S_{\rm m}}{\delta \phi} \right] + \frac{1}{2M^2} \left[ 4 V (\phi) + \rho - 3 p \right] \right) F_{\mu \nu} \tilde F^{\mu \nu} \, , \label{S_h1}
\end{align}
Working in the Coulomb gauge in the spatially flat Friedmann metric, we get 
\begin{equation}\label{S1h}
S_{\rm a1} = 12 \chi_1 \int d^4x\, e^{\phi / M} \left( \frac{1}{M} \left[ V' (\phi) - \frac{\delta S_{\rm m}}{\delta \phi} \right] + \frac{1}{2M^2} \left[ 4 V (\phi) + \rho - 3 p \right] \right) \epsilon_{ijk} A_i^{T\prime} \partial_j A_k^T \, .
\end{equation}

\subsection{Evolution of magnetic field}

Assuming that the axial coupling dominates in the action, and decomposing the Fourier modes of the transverse part $A^T_i$ in the helicity basis $A^T_i (\eta, \bk) = \mathcal{A}_+ \varepsilon_{i, \bk}^+ + \mathcal{A}_- \varepsilon^-_{i, \bk} \,$, one arrives at the following equation for the helicity modes:
\begin{equation}\label{A_h}
\mathcal{A}''_h + \left[ k^2 + h k w' \right]\mathcal{A}_h = 0 \, ,
\end{equation}
where $h=\pm 1$ denotes the helicity, and
\begin{equation} \label{w}
w =  12 \chi_1 e^{\phi / M} \left(\frac{1}{M} \left[ V' (\phi) - \frac{\delta S_{\rm m}}{\delta \phi} \right] + \frac{1}{2M^2} \left[ 4 V (\phi) + \rho - 3 p \right] \right) \, .
\end{equation}
Using the inflationary slow-roll conditions (\ref{slow-roll}) and expression (\ref{comb}), we have
\begin{equation}
w = 12 \chi_1 m^2 \left( e^{\phi / M} - 1 \right) \, ,
\end{equation}
\begin{equation}
w' \approx \frac{16 \chi_1 m^2}{\eta}\left[ 1 + \mathcal{O} \left( \epsilon_V, \eta_V \right) \right] \, .
\end{equation}
Then, equation (\ref{A_h}) takes the form
\begin{equation}
\mathcal{A}''_h + \left[ k^2 + 16 \chi_1 m^2 \, \frac{h k}{\eta} \right]\mathcal{A}_h = 0 \, .
\end{equation}
This equation should be supplemented by the vacuum initial condition
\begin{align}
\mathcal{A}_h(\eta, k) \simeq \mathcal{A}_{\text{vac}}(\eta, k) = (2k)^{-1/2}e^{- \ri k\eta} \quad \text{for} \quad - k \eta \gg 1 \, .
\end{align}

It is convenient to introduce a new dimensionless variable $x=-k\eta$, so that inflation starts when $x \gg 1$ and ends as $x \rightarrow + 0$.  We then have 
\begin{equation}\label{A(x)}
\mathcal{A}''_h (x) + \left[ 1 - \frac{h \xi}{x} \right]\mathcal{A}_h (x) = 0 \, , 
\end{equation}
\begin{equation}
\mathcal{A}_h(x) \rightarrow (2k)^{-1/2}e^{\ri x} \quad \text{as} \quad x \to \infty \, .
\end{equation}
Here, $\xi  \equiv  16 \chi_1 m^2$ is a dimensionless parameter that controls the amplification of the helical magnetic field amplitude during inflation. 

Solutions to equation (\ref{A(x)}) are expressed in terms of the regular and irregular Coulomb functions $F_0$ and $G_0$, and the combination that fits the initial conditions is the following:
\begin{equation}
\mathcal{A}_h(x) = (2k)^{-1/2} \left[ G_0 (h \xi / 2, x) + \ri F_0 (h \xi / 2, x) \right] \, .
\end{equation}
This solution was obtained in \cite{Anber:2006xt} in the context of $N$-flation and in \cite{Durrer:2010mq} for the case of constant axial coupling in the general single-field inflationary model. Using the asymptotic expressions for the functions $F_0$ and $G_0$ as $x \rightarrow 0$ \cite{Durrer:2010mq}, one arrives at the following expression for the field modes at the end of inflation:
\begin{equation}\label{A_final}
\mathcal{A}_h(x) \underset{x \to 0_+}{\longrightarrow} (2k)^{-1/2}  \left[\frac{\exp(h \pi \xi / 2)\sinh(\pi \xi / 2)}{\pi \xi / 2} \right]^{1/2} \, .
\end{equation}
A divergence-free statistically homogeneous and isotropic magnetic field has the following general Fourier representation of the two-point correlation function:
\begin{equation}
\left\langle B_i(\eta, \bk) B^*_j(\eta, \bk') \right\rangle = \frac{(2\pi)^3}{2} \, \delta(\bk - \bk') \left[ \left(\delta_{ij} - \hat{k}_i \hat{k}_j \right) P_{S} (\eta, k) - i \epsilon_{ijl} \hat{k}_l P_{A} (\eta, k) \right] \, ,
\end{equation}
where $\hat{k}_i \equiv k_i/k$. Inflationary cosmology equates this late-time stochastic power spectra to the vacuum expectation values of quantum fields, which allows one to express the functions $P_{S} (k)$ and $P_{A} (k)$ in terms of the field modes:
\begin{equation}
P_{S/A} (\eta, k) = k^2 \left( \left| \mathcal{A}_+ (\eta, k) \right|^2 \pm \left| \mathcal{A}_- (\eta, k) \right|^2 \right) \, .
\end{equation}
Using the asymptotic expression (\ref{A_final}), we get 
\begin{equation}
P_{S} (k) = k \frac{\sinh (\pi \xi)}{\pi \xi} \, , \qquad P_{A} (k) = k \frac{\cosh (\pi \xi) - 1}{\pi \xi}\, .
\end{equation}

Although the scaling of the power spectra with respect to $k$ is the same as in the vacuum case, they get multiplied by an amplification prefactor which depends on the value of $\xi$. Following the general ideology of effective field theories, one might argue that the value of the axial coupling $\chi_1$ should be of the order $M_{\rm UV}^{-2}$, where $M_{\text{UV}}$ is the mass scale (or one might say cutoff) that characterizes the UV-complete theory, which in the Starobinsky model could be comparable to the inflaton mass $m \simeq 10^{-5} M$, so that $\chi_1 m^2 \sim 1$. In this case, without too much fine-tuning, one can get a considerable amplification if the condition $\pi \xi = 16 \pi \chi_1 m^2 \gtrsim 100$ is satisfied. Then the amplification factor could be as large as $\left(e^{\pi \xi}/{\pi \xi}\right)^{1/2}\sim 10^{25}$ which would allow to achieve minimal $10^{-20}$~G required for subsequent dynamo amplification. However, back-reaction considerations appear to constrain the amplification factor severely.  We investigate this issue in the next subsection.

\subsection{Back-reaction}

The first line in (\ref{S_h1}) gives the following contribution to the stress--energy tensor:
\begin{align}
\Delta T_{\mu\nu} \equiv  \frac{2}{\sqrt{- g}} \frac{\delta S_{\rm a1}}{\delta g^{\mu\nu}}	= \chi_1 e^{\phi/M} \left[ 2 \nabla_\mu \nabla_\nu \cF  - 2 R_{\mu\nu} \cF + \frac{10}{M} \nabla_{(\mu} \phi \nabla_{\nu)} \cF + \frac{2}{M} \nabla_\mu \nabla_\nu \phi \cF  \right. \nonumber \\  \left. {} + \frac{11}{M^2} \nabla_\mu \phi \nabla_\nu \phi \cF  - g_{\mu\nu} \left(2 \Box \cF + \frac{5}{M} \Box \phi \cF + \frac{7}{M} \nabla_\alpha \phi \nabla^\alpha \cF + \frac{5}{M^2} \nabla_\alpha \phi \nabla^\alpha \phi  \cF \right)    \right]	\, ,
\end{align}
where $\cF \equiv F_{\mu \nu} \tilde F^{\mu \nu}$.  Using the Einstein equations and the equation of motion for the scalar field, we transform this to
\begin{align}
	\Delta T_{\mu\nu} = \chi_1 e^{\phi/M} \left[ 2 \nabla_\mu \nabla_\nu \cF  + \frac{10}{M} \nabla_{(\mu} \phi \nabla_{\nu)} \cF + \frac{2}{M} \nabla_\mu \nabla_\nu \phi \cF  + \frac{14}{M^2} \nabla_\mu \phi \nabla_\nu \phi \cF \right. \nonumber \\  \left. {} - g_{\mu\nu} \left(2 \Box \cF - \frac{5}{M} V' (\phi) \cF + \frac{7}{M} \nabla_\alpha \phi \nabla^\alpha \cF + \frac{8}{M^2} \nabla_\alpha \phi \nabla^\alpha \phi \cF - \frac{9}{M^2} V (\phi) \cF \right)    \right]	\, .	
\end{align}
The contribution to the energy density is then
\begin{align}
	\Delta \rho = \Delta T_0{}^0 = 3 \chi_1 e^{\phi/M} \left[ - 2 H \dot \cF + \frac{1}{M} \dot \phi \dot \cF - \frac{2}{M} H \dot \phi \cF + \frac{1}{M} V' (\phi) \cF \right. \nonumber \\ \left. {} + \frac{2}{M^2} \dot \phi^2 \cF + \frac{3}{M^2} V (\phi) \cF    \right] \, .
\end{align}
Leaving only the terms dominant during inflation, we can simplify this expression to
\begin{equation}\label{rho}
\Delta \rho = 6 \chi_1 e^{\phi/M} \left( 3 H^2 \cF - H \dot \cF \right) \, .	
\end{equation}

We will need the following expressions for the relevant quantities in terms of the mode functions:
\begin{align}
\rho_B &= a^{-4} \int_0^{k_{\rm f}} \frac{dk \, k^4}{(2\pi)^2} \left[ \left| \mathcal{A}_+ (\eta, k) \right|^2 + \left| \mathcal{A}_ -  (\eta, k) \right|^2 \right] \, , \\
\rho_E &= a^{-4} \int_0^{k_{\rm f}} \frac{dk \, k^2}{(2\pi)^2} \left[ \left| \mathcal{A}^{\prime}_+ (\eta, k) \right|^2 + \left| \mathcal{A}^{\prime}_ - (\eta, k) \right|^2 \right] \, , \\
\cF = F_{\mu \nu} \tilde F^{\mu \nu} &= 4a^{-4} \int_0^{k_{\rm f}} \frac{dk \, k^3}{(2\pi)^2} \left[ \left| \mathcal{A}_ + (\eta, k) \right|^2 - \left| \mathcal{A}_ - (\eta, k) \right|^2 \right]^{\prime} \, . 
\end{align}
Here, $\rho_B$ and $\rho_E$ are energy densities of the magnetic and electric field, respectively,  stemming from the free action (\ref{Maxwell}), and $k_{\rm f} = (aH)_{\rm f}$ is the wavenumber of the last mode that crosses the horizon during inflation. Denote the amplification factor appearing in (\ref{A_final}) by $W_{h}(\xi)$:
\begin{equation}
W_{h}(\xi) = \left[\frac{\exp(h \pi \xi / 2)\sinh(\pi \xi / 2)}{\pi \xi / 2} \right]^{1/2} \, .
\end{equation}
We are interested in the case where $W^2_{h}(\xi) \equiv W^2(\xi) \gg 1$ for one of the helicities; for the opposite helicity, this quantity will then be of order $1/\pi\xi$. Using the asymptotic expressions
\begin{equation}
\mathcal{A}_h(\eta, k) \underset{\eta \to 0_-}{\longrightarrow} \left( 2k \right)^{-1/2} W_{h}(\xi) \, ,
\end{equation}
\begin{equation}
\mathcal{A}^{\prime}_h(\eta, k) \underset{\eta \to 0_-}{\longrightarrow} - \left( k/2 \right)^{1/2} \left[h\xi W_{h}(\xi) \log \left( h \xi k|\eta| \right) + \ri W^{-1}_h(\xi) \right] \, ,
\end{equation}
which are valid for  $k|\eta| \ll 1/4\xi $ \cite{Durrer:2010mq}, as $\eta \to 0_-$ we get
\begin{align}
\rho_B &= \frac{\sinh (\pi\xi)}{\pi \xi}\frac{H_{\rm f}^4}{4 (2 \pi)^2} \simeq \frac{W^2(\xi) H_{\rm f}^4}{8 (2 \pi)^2} \, , \\
\rho_E &= \mathcal{O} (1)\, \frac{\xi^2 \, W^2(\xi) H_{\rm f}^4}{8 (2 \pi)^2} \, ,
\end{align}
where $H_{\rm f} \simeq m/4$ is the Hubble parameter at the end of inflation. To compute contribution (\ref{rho}), we find
\begin{align}
\cF &= F_{\mu \nu} \tilde F^{\mu \nu} =  \mathcal{O} (1)\, \frac{\xi \, W^2(\xi) H_{\rm f}^4}{(2 \pi)^2} \, , \label{cF} \\
\dot \cF &=  \mathcal{O} (\xi^2 , \xi)\, \frac{W^2(\xi) H_{\rm f}^5}{(2 \pi)^2}  \, , \\
\Delta \rho &= \mathcal{O} (\xi^3 , \xi^2)\, \frac{W^2(\xi) H_{\rm f}^4}{8 (2 \pi)^2} \, ,
\end{align}
where we have dropped numerical coefficients of order unity and corrections, logarithmic with respect to $\xi$.  Combining all contributions, we obtain
\begin{equation}
\rho_B + \rho_E +\Delta \rho = \mathcal{O} (\xi^3 , \xi^2, 1)\, \frac{W^2(\xi) H_{\rm f}^4}{8 (2 \pi)^2} \, .
\end{equation}

In order for the generated field not to spoil inflation, this resulting contribution to the energy density should not exceed that of the inflaton:
\begin{equation}
\rho_B + \rho_E +\Delta \rho \, \lesssim \, \rho_{\phi} = \frac12 \dot{\phi}^2 + V(\phi) \, .
\end{equation}
This allows to make an upper bound on the amplification factor:
\begin{equation}
\mathcal{O} (\xi^3 , \xi^2, 1) \, W^2(\xi) \, \lesssim \, 10^4 \left(\frac{M}{m}\right)^2 \sim 10^{14} \, ,
\end{equation}
leading to
\begin{equation}\label{W}
W(\xi) \lesssim 4 \times 10^5 \, , \qquad \chi_1 m^2 \lesssim 0.6 \, .
\end{equation}
Similar estimates were obtained in a slightly different context in \cite{Durrer:2010mq}.

Another restriction on the amplification factor comes from the back-reaction on the dynamics of the inflaton field. The couplings in (\ref{S_h1}) lead to the following modified equation of motion for the scalar field:
\begin{equation} 
\Box \phi + V'(\phi) + \frac{\chi_1}{M} e^{\phi / M} \left[R +  \frac{9}{M} \Box \phi + \frac{9}{2M^2}(\nabla\phi)^2 + \frac{9}{M}  \nabla_{\mu}\phi \nabla^{\mu} + 3 \Box \right] \mathcal{F}  = 0 \, .
\end{equation}
By using first-order equations of motion, this equation is reduced to
\begin{equation} 
\Box \phi + V'(\phi) - \frac{3\chi_1}{M} e^{\phi / M} \left[ \frac{2}{M^2} V(\phi) + \frac{3}{M} V^{\prime}(\phi) - \frac{2}{M^2}(\nabla\phi)^2 - \frac{3}{M}  \nabla_{\mu}\phi \nabla^{\mu} - \Box \right] \mathcal{F}  = 0 \, ,
\end{equation}
and, leaving only the terms dominant during inflation, we have
\begin{equation} 
\Box \phi + V'(\phi) + \frac{3\chi_1}{M} e^{\phi / M} \left[ \ddot{\mathcal{F}} + 3H\dot{\mathcal{F}} - m^2 \mathcal{F} \right]  = 0 \, .
\end{equation}

In order for the additional exponential term not to spoil the usual slow-roll inflationary regime, it must be considerably smaller than $V'(\phi)$. Using the fact that during inflation $\dot{\mathcal{F}} \simeq H \mathcal{F}$ and plugging in the asymptotic expression (\ref{cF}), we obtain the condition
\begin{equation}
\mathcal{O} (\xi^2) \, W^2(\xi) \, \lesssim \, \left(\frac{M}{m}\right)^2 \, ,
\end{equation}
which is essentially the same bound as (\ref{W}).

\subsection{Preheating}

At the preheating stage, since the created matter is relativistic and satisfies $\rho - 3 p = 0$, and neglecting the term ${\delta S_{\rm m}}/{\delta \phi}$ in (\ref{w}), we have, to linear order in $\phi / M$,
\begin{equation}
w \approx 	12 \chi_1 m^2 \frac{\phi}{M} \, .
\end{equation}
The free scalar field evolves as $\phi = \phi_a \sin \left( m \int a d \eta \right)$ with slowly decreasing amplitude $\phi_a \propto a^{-3/2}$.  Equation (\ref{A_h}) in this case becomes, to the leading terms,
\begin{equation}
\mathcal{A}''_h + \left[ k^2 + 12 h k \chi_1 m^2 \frac{\phi_a}{M} a m \cos \left( m \int a d \eta \right) \right]\mathcal{A}_h = 0 \, .	
\end{equation}
The narrow-resonance condition for this equation in the lowest resonance zone for both helicities is the same:
\begin{equation}
\left| \frac{4 k^2}{a^2 m^2} - 1 \right| < \frac{6 k}{am} 	\chi_1 m^2 \frac{\phi_a}{M}\, .
\end{equation}
For $\chi_1 m^2 \lesssim 1$, the resonance occurs in a narrow band around $k / a \approx m / 2$, leading to the usual resonant creation of high-frequency photons.  

\section{Both types of coupling in the action}
\label{sec:both}

Suppose that both types of couplings are present in the action, and consider the case when $S_1$ and $S_{\rm a1}$ are important.  First of all, we note that one can add an arbitrary constant to $w$ defined in (\ref{S1h}) and (\ref{w}), since this will only add a surface term to the action.  One then can adjust this constant in such a way that $I \equiv I_+$ and new $w$ become proportional to each other.  After that, the total action for the electromagnetic field takes the form
\begin{equation}
S = \frac12 \int d^4 x\, I^2 \left( A_i^{T\prime} A_i^{T\prime} + A_i^T \Delta A_i^T + \frac{4 \chi_1}{\kappa_1} \epsilon_{ijk} A_i^{T\prime} \partial_j A_k^T \right) \, .
\end{equation}
The theory with such an action and with $I \propto a^n$ was under consideration in \cite{Caprini:2014mja}.  In our case, we have [see (\ref{Ipm1})]
\begin{equation}
I^2 = 1 + 12 \kappa_1 m^2 \left( \frac43 \log \frac{a_{\rm f}}{a} + 1 \right) \, .
\end{equation}
Then,  proceeding to variables (\ref{vi}) with $I_+ = I$ and expanding in the helical basis $v_i (\eta, \bk) = v_+ \varepsilon_{i, \bk}^+ + v_- \varepsilon^-_{i, \bk} \,$, we obtain the following equation for the variables $v_h$, $h = \pm$:
\begin{equation}\label{vhev}
v''_h + \left( k^2  + 4 h k \frac{\chi_1 I'}{\kappa_1 I} - \frac{I''}{I} \right) v_h = 0 \, .
\end{equation}

The last term in the parentheses is growing faster than the second one during inflation.  So, the only non-trivial scenario is that initially the second term dominates over the last one, and the evolution is described by the equations of the previous section, in which case $v_h = C (\xi) I$, where $C (\xi)$ is given by the right-hand side of (\ref{A_final}), and afterwards, when the last term in (\ref{vhev}) starts dominating, the evolution $v_h \propto I$ remains to be valid.  Thus, we obtain the result close to (\ref{A_final}) even in this more complicated scenario. 

\section{Discussion}
\label{sec:discuss}

In this work, we investigated the possibility of primordial magnetogenesis in one of the curvature-based inflationary models\,---\,the Starobinsky model. The idea is that, just as one can consider the $R^2$ term as a leading quantum correction in a low-energy effective field theory expansion, one can also include in this expansion all possible couplings between the electromagnetic field and curvature. After proceeding to the Einstein frame by conformal transformation of the metric, these terms generate nontrivial couplings of the electromagnetic field and the inflaton, with the form of interaction completely determined by the original model without further arbitrariness.  This idea can be realized also in other inflationary models with non-minimal couplings to curvature, such as the Higgs inflation \cite{Bezrukov:2007ep}. 

We studied the most general terms of the smallest dimension six in the effective action, which can be regarded as the lowest-order loop corrections to electromagnetism in curved spacetime. In a general case, these couplings involve both parity-preserving and axial parts, with the axial part leading to helicity-dependent evolution of electromagnetic field. Considering electromagnetic field as a test field that does not influence the cosmological expansion or the dynamics of the inflaton, we analyzed the evolution equations for the mode functions during inflationary regime and preheating. In the parity-preserving case, amplification of the magnetic field is negligibly small for the values of couplings allowed by the requirement of stability of electromagnetic field during and after inflation.  In the axial case, one can get a significant amplification in a natural range of parameters. However, back-reaction considerations, based on the requirement that the generated electromagnetic field should not spoil the usual inflationary evolution, lead to a strong constraint $W \lesssim M/m \sim 10^5$ on the amplification factor of the vacuum fluctuations, resulting in the spectrum $B_\lambda \lesssim 10^{-52} \left({\rm Mpc}/\lambda \right)^2$~G today all the way down to the cosmic diffusion scale $\lambda_{\rm diff} \sim 1\,{\rm AU} \approx 5 \times 10^{-6}\,{\rm pc}$ \cite{Grasso:2000wj}, at which one would have $B \lesssim 10^{-30}$~G\@. This is apparently not enough to explain the observed large-scale magnetic fields. Such a strong constraint can be traced back to the fact that, in all cases, due to the flatness of the Starobinsky potential, the coupling functions change very slowly during inlfation, so that the form of the power spectrum of electromagnetic field in the amplification domain is not essentially modified with respect to the vacuum case. 

As we have established, the back-reaction of the generated electromagnetic field in this model is two-fold.  Firstly, the stress--energy contribution from the new coupling terms in the action modifies the dynamics of the universe expansion, threatening to halt inflation.  Secondly, the same new terms in the action modify the dynamics of the inflaton.  Both effects turned out to be of the same level of importance for the estimation of back-reaction effects in the model under consideration.  The specific flatness of the scalar-field potential in models of this type leads to an interesting question about possible existence of a self-consistent regime whose inflationary property is modified but not completely destroyed by the presence of generated electromagnetic field, and which might be free from the back-reaction constraint.  We leave this as a subject of future investigation.

\section*{Acknowledgments}

The work of O.~S. was supported by the ICTP -- SEENET-MTP project PRJ-09 ``Cosmology and Strings'' and project NT-03 ``Cosmology -- Classical and Quantum Challenges.'' The work of Y.~S. was supported by the National Academy of Sciences of Ukraine (project No.~0116U003191) and by grant 6F of the Department of Targeted Training of the Taras Shevchenko National University of Kiev under the National Academy of Sciences of Ukraine.

\end{document}